\documentclass[aps,prd,superscriptaddress,longbibliography,amssymb,twocolumn]{revtex4-1}

\usepackage[utf8]{inputenc}
\usepackage[pdftex]{graphicx}
\usepackage{amsmath}
\usepackage[amssymb,Gray]{SIunits}
\usepackage{textcomp}
\usepackage{physics}
\usepackage{amssymb}
\usepackage{mathtools}
\usepackage{tikz}
\usepackage{hyperref}
\usepackage{float}
\usepackage[most]{tcolorbox}

\usepackage{graphicx} % Required for inserting images
\usepackage{comment}

\usepackage{xcolor}
\newcommand{\mn}{\mu \nu}
\newcommand{\h}[1]{\hat{#1}}

\renewcommand{\vec}[1]{\mathrm{\mathbf{#1}}}

\begin{document}

%\preprint{}

\title{Environmental gravitational decoherence with higher derivative theory}


%\author{}
%\email[]{}
%\thanks{}
%\affiliation{}
%
\author{Linda M. van Manen}
\email[]{linda.van.manen@uni-jena.de}
\affiliation{Institute for Theoretical Physics, Friedrich Schiller University Jena, Fr\"obelstieg 1, 07743 Jena, Germany}


\date{December 30, 2024}

% PRL: abstract max. 600 characters
\begin{abstract}
We discuss the decoherence in a quantum system induced by interaction with gravitational degrees of freedom that are part of a higher derivative theory. The deformation of a mass distribution due to gravitational waves acquires naturally a mass quadrupole moment. This adds higher derivative dynamics of the quadrupole moment to the unitary evolution of the system, where the quadrupole moment oscillates with the gravitational frequencies following a higher derivative theory. The consequence of higher derivatives in the dynamics is that the system is described by four canonical variables. This departure from the usual particle position and momentum operators gives an entirely different interpretation of the decoherence basis. This model focuses on the open dynamics of the quadrupole moment, rather than on individual particles. As such, a short example is given to utilize quadrupole measurements to probe gravitational decoherence and noise. We first derive a Langevin equation for a lower derivative model and show how higher derivatives naturally emerge on the boundary. A quantum master equation is derived for the emerging quadrupole moment, considering that the environment is a higher derivative theory of gravity. 
\end{abstract}

%\keywords{}


\maketitle

\section{Introduction}
Gravitational decoherence in quantum systems due to interaction with a gravitational field has been of interest in recent years. It is one of the few methods to gain experimental data about gravity in a non-relativistic, weak field regime \cite{SCHLOSSHAUER20191, marshall2003, parikh, Yu2020, Suzuki_2015}. These gravitational interactions are universal and gravitational decoherence is present in all experiments, although the extremely small scale makes it challenging to observe any quantum features. With the detection of gravitational waves \cite{abbot2015} a new era of experimental data presented itself. It has provided us with valuable astronomical data and opened new possibilities to obtain data on the fundamental nature of gravity \cite{Yu2020, parikh}. 

In light of this, work has been done in recent years on gravitational radiation, interaction with external gravitational waves, and noise signatures of graviton emission/absorption in gravitational wave detectors. Such have Suzuki et al. \cite{Suzuki_2015} evaluated environmental gravitational decoherence in an interference experiment, and Parikh et. al \cite{parikh} showed that a gravitational squeezed state could in principle be measured in space-borne gravitational wave detectors such as LIGO \cite{amaroseoane2017laser}. Furthermore,  Oniga et al. \cite{Oniga2017} have evaluated decoherence from an ensemble of particles. The gravitational radiation in the classical limit was related to the radiation quadrupole formula.  In a recent article \cite{toros2024}, Toros et al. studied graviton emission from a linear quadrupole moment associated with a harmonically trapped particle. The issues that are generically, although not exclusively, being addressed in this research are the decoherence rate, the associated basis, and how to distinguish gravitational radiation from other radiation sources. It was stated by Anastopoulos et al. \cite{Anastopoulos_2013} that a specific feature of gravitational decoherence is that decoherence appears in the momentum (energy) basis, and not necessarily in the position basis. While Asprea et al. \cite{asprea2021} showed the decoherence in position, momentum, and energy basis in their appropriate limits, and explained the differences.

In this work, we evaluate a free multi-particle system exclusively interacting with external quantized gravitational waves. The focus here lies on the influence of the gravitational waves on the unitary part of the system, rather than on radiation. Some known features are the frequency shift, analogous to the Lamb shift, and a mass correction \cite{barone1991}. What is unique to gravity is a fourth derivative boundary term. This distinguishes interaction with gravity from interaction with other media and has larger implications than the lower derivative correction \cite{simon}. It changes the unitary part of the system to a higher derivative theory, which leads to a significantly different theory, even for weakly coupled systems. It suggests that the unitary evolution of the systems gains a larger set of solutions than it had before the interaction. Furthermore, the position and conjugate momenta operators need to be re-evaluated, with consequences for the decoherence basis.

Higher derivative theory means a theory with a Lagrangian containing derivatives higher than the first derivative and was first described by Ostrogradsky \cite{ostrogadsky1850, woodard2015}. Higher derivatives appear naturally in many places of physics (see \cite{Podolsky1942, weyl, lovelock1971}) and have been of interest for (quantum) gravity \cite{Gràcia1991, LIU2013413, Brandenberger, Charmousis2009}. The most notable is the quadratic Riemann tensor in the classical gravitational action, which has been shown \cite{osti_4774739, Stelle77} to re-normalize divergences from quantum corrections that arise when interacting with matter fields. These theories have been considered nonphysical by some due to the Ostrogradsky instability \cite{kaparulin2014,woodard2015,simon}, i.e., instability due to a lack of lower bound in the Hamiltonian, although there is quite some evidence that systems with unbounded Hamiltonians can be perfectly stable \cite{ErrastiDiez:2024hfq, Deffayet2023, Ganz_2021}.

In this article, we focus on a simple implementation of Ostrogradsky's higher derivative theory, where the gravitational Hamiltonian, as well as the systems quadrupole Hamiltonian, are a harmonic oscillator plus a second derivative in the ``position" variable. With quadrupole Hamiltonian we mean the Hamiltonian that describes the dynamics of the quadrupole, rather than the individual particles. The higher derivatives added to the unitary part of the system model do not have any instability issues. This is because the higher derivatives in the Hamiltonian cancel the higher derivative boundary terms in the equation of motion, and thus do not contribute to run-away solutions.

The focus on the quadrupole dynamics in this article is not only due to the simplistic manner the system can be described when described in terms of its quadrupole dynamics induced by gravitational waves. We shortly discuss the idea to investigate the effect of gravitational waves on the hyperfine structure. The idea is that not only the mass quadrupole moment is affected, but the electric quadrupole moment as well, which can be measured via, e.g., the M\"{o}ssbauer effect \cite{Padilla-Rodal_2011, JOHNSON1962129}. One could, for instance, consider the mass density of a nucleus, which is being deformed by gravitational waves. The deformation of mass equals a deformation of charge, thus creating an electric quadrupole moment. The measurements are extremely precise and give sharp lines of the hyperfine structure of a nucleus, making it sensitive to extremely small changes.

We close this article by deriving the quantum master equation for the quadrupole moment. We find a similar decoherence rate than reported earlier in literature \cite{toros2024}, although it shows a very different definition for the basis in which decoherence is expected.

\section{The model}

We first review a microscopic lower-derivative model for a system interacting with a gravitational field acting as the environment. We derive the stochastic equation of motion, following the method described in \cite{Borghini2012,lampo}. Finally, we discuss the relation with the known classical radiation equation from general relativity \cite{maggiore}, and show how the higher derivative boundary term emerges.

\subsection{The environment}
The gravitational field is considered a weakly perturbed spacetime, described by the linear metric

$$g_{\mn} = \eta_{\mn} + h_{\mn}, \quad h_{\mn} << 1.$$
The interaction between the particles and their gravitational fields is neglected. Hence we can consider the vacuum solution of the Einstein equations, and we apply the transverse-traceless (TT) gauge \cite{maggiore}. This results in  gravitational waves, conveniently discretized by placing the field in a box with length $L$:

\begin{align}
    h_{ij}^{TT}(\vec{x},t) &= \frac{1}{L}\sum_{k, \lambda =+,\times}^{\Omega}  \;q_{k,\lambda}(t) e^{i \vec{k} \cdot \vec{x}}\; \epsilon_{ij}^{\lambda,\;TT}(k),
    \label{metric_h}
\end{align}
with $q_{k}(t) = q_0 e^{i\omega_k t},$  $\epsilon_{ij}^{\lambda,\;TT}(k)$ is the polarization tensor, and where $\Omega$ is a high frequency cut-off.. For clarity, we will drop the superscript $\lambda$ from here on. The resulting Einstein-Hilbert action is given as a set of harmonic oscillators

\begin{align}
    S_{EH} =& \frac{c^3}{64 \pi G}\int d^4x \; \partial_{\mu} h^{ij}_{TT} \partial^{\mu} h_{ij}^{TT}\\
    =& -\frac{c^2 L}{32 \pi G}\int dt  \sum_{k}  \Dot{q}_{k}^2(t) - \omega^2_k q_{k}^2(t)
\end{align}
where it has been used that
\begin{align}
        \int d^3x \; e^{i(k+k')x} &= L^3 \; \delta^{(3)}(k'+ k).
\end{align}
Furthermore, $|k| = \frac{\omega_k}{c}$, $x_0 = - c t$, and the polarization tensor is normalized as \cite{maggiore}

$$\epsilon_{ij}^{\lambda}(k)\;  \epsilon^{ij\,\lambda' }(k)  = 2 \delta^{\lambda, \lambda'}.$$

\subsection{The Brownian particles}
The system consists of $n$ masses with equal mass $m$ on a geodesic parameterized by $x_n(t_n)$. The particles are moving non-relativistically and are interacting with the bath of mutually independent harmonic oscillators.

The interaction can be evaluated at certain length scales related to the energy limit of interest. This scale, however, will be considered much smaller than the wavelength of the external gravitational radiation (i.e. the analog of the electric dipole approximation). Hence the particles are separated by a distance $|\xi|$, much smaller than the cut-off $\Omega$ given in (\ref{metric_h}), which determines the high- or low-frequency regime that is being considered. With this approximation we have $e^{i\vec{k}\vec{x}} = 1 + \dots$,  such that $h_{\mn}(x_1) \approx h_{\mn}(x_2)$, and  $t_2 \approx t_1$, for all $n$.

For our coordinate frame, we choose the simplest laboratory frame used by experimentalists to describe its apparatus, which is a drag-free satellite \cite{maggiore}. This leads to
\begin{widetext}
$$ds^2 = - c^2 dt^2 \left[1 + R_{0i0j} x^i x^j \right] - 2 c dt dx^i \left[\frac{2}{3} R_{0jik} x^jx^k \right] + dx^idx^j \left[\delta_{ij} - \frac{1}{3} R_{ikjl} x^kx^l \right],$$
\end{widetext}
where the Riemann tensor is evaluated at the points $x_n(t)$, and we utilize the gauge invariance of the Riemann tensor, $$R_{i0k0} = R_{i0k0}^{TT} = \frac{1}{2 c^2} \Ddot{h}^{TT}_{ij}.$$

The action for $n$ testparticles  parametrized by $x_n(t)$ is given by
\begin{align}
    S_p =& -\frac{m}{2}  \sum_n  \int dt \left[ \delta_{ij} \;\Dot{x}^{i}_n\Dot{x}^{j}_n + R_{0i0j} x^i_n x^j_n \right]\nonumber \\
    =& - \sum_{k,n} \int dt \; \frac{m}{2}  \;\Dot{x}^{i}_n\Dot{x}^{j}_n \delta_{ij} + \frac{\omega^2_k m}{4L} q_k(t) \; \epsilon_{ij} x^i_n x^j_n,
\end{align}
where the rest energy is neglected. The second term in the action is the interaction part, where we can write the sum over $x_n$ as an integral 
\begin{align}
    \sum_{n}  x^i_n x^j_n &=  \int dx^3  \; x^i x^j \;\delta^3 (x - x_n),
\end{align}
and write the mass density as $\rho(x,t) = m \; \delta^3 (x - x_n(t))$. The  interaction can now be recognized as an interaction between the acceleration of the environmental degree of freedom $\ddot{q}(t) =  \omega^2_k q(t)$ and the mass quadrupole moment, defined as 

\begin{align}
    Q_{ij}(t) \equiv \int d^3x \; \rho(x,t) x_i x_j +\frac{r}{3}\delta_{ij},
    \label{quad}
\end{align}
where the term proportional to the delta function is zero in the TT gauge.

The full action is then given by
 \begin{widetext}
\begin{align}
    S = &- \int dt \sum_{k}  \frac{\mu}{2} \left[\Dot{q}_{k}^2(t) - \omega_k^2 \,q_{k}^2(t)\right]  - \sum_{n} \int dt \; \frac{m}{2}  \;\Dot{\vec{x}}_n\Dot{\vec{x}}_n - \sum_{k} \int dt \; \frac{\omega^2_k }{4L} q_k(t) \; Q^{ij}(t)  \epsilon_{ij},
\end{align}
\end{widetext}
with $\mu \equiv \frac{L c^2 }{16 \pi G}$ being the ``mass" of the environment.

\subsection{The Hamiltonian}
The canonical momenta $p$  and  $\pi$ for the particle and bath, respectively,  are given by

\begin{align}
    \vec{p}_n =& \frac{\partial L}{\partial \Dot{\vec{x}}_n} = m \Dot{\vec{x}}_n,\\
    \pi_k =&   \frac{\partial L}{\partial \Dot{q}_k} =   \mu \;\Dot{q}_{k},
\end{align}

which leads to the classical Hamiltonian

\begin{align}
    H =& \sum_{k,n}\Big[\frac{\vec{p}_n^2}{2m} + \frac{\omega_k^2  }{4L}\;q_k \; Q^{ij} \epsilon_{ij}\Big] + \frac{\pi^2_k}{2 \mu} + \frac{\mu}{2} \omega^2_k q_k^2.
\end{align}
 The first two terms in brackets are the system's kinetic energy and the interaction with the gravitational field. The last two terms we recognize as the harmonic oscillator describing the environment. 
 \subsection{The quantum description}
For the quantum Hamiltonian, we introduce the ladder operators for the environment, 
\begin{align}
    \h{q}_k(t) &= \sqrt{\frac{\hbar}{2 \mu \omega_k}} (\hat{a}_k^{\dagger}(t)+\hat{a}_k(t)),\\
    \h{\pi}_k(t) &= i \sqrt{\frac{\hbar \mu \omega_k}{2}}(\hat{a}^{\dagger}_k(t)-\hat{a}_k(t)),
\end{align}
such that the Hamiltonian can be rewritten as
\begin{align}
    \hat{H} =& \hat{H}_s  + \sum_k  \hbar \omega_k \left(\hat{a}_k^{\dagger}(t) \hat{a}_k(t)+\frac{1}{2}\right)\\
    &+\sum_{k,n} \sqrt{\frac{\hbar \omega_k^3 m^2}{32 L^2 \mu}} \; (\hat{a}_k^{\dagger}(t)+\hat{a}_k(t))\;\hat{x}_n^i \; \hat{x}_n^j \epsilon_{ij},
\end{align}
with $\hat{H}_s = \sum_n \frac{\hat{\vec{p}}_n^2}{2m}$. We derive the dynamics for the individual particles first and shift our focus to the dynamics of the quadrupole moment afterward.

\section{Dynamics}
The corresponding equations of motion are the Heisenberg-Langevin equations for the particle,

\begin{align}
    \frac{d \hat{x}_n^i}{dt} & = \frac{\hat{p}_n^i}{m}\\
    \frac{d \hat{p}_{nj}}{dt} & = - \sum_k \sqrt{\frac{\hbar \omega_k^3 m^2}{8 \mu L^2 }}(\hat{a}_k(t)+\hat{a}^{\dagger}_k(t)) \hat{x}_n^i \epsilon_{ij}, \label{eq_p}\\
 \label{eq_pi}
\end{align}
and the Heisenberg equation for $\hat{a}_k(t)$ (and $\hat{a}_k^{\dagger}(t)$) is given by

\begin{align}
    \frac{d \hat{a}_k(t)}{dt} &= \frac{i}{\hbar}[\hat{a}_k(t), \hat{H}]\\
    &= - i \omega_k \hat{a}_k(t) - i \sum_m \sqrt{\frac{\omega_k^3 m^2}{32 \hbar L^2 \mu}} \;\epsilon_{ij}\; \hat{x}^i_m(t) \hat{x}^j_m(t) ,
\end{align}
which has a solution

\begin{widetext}
\begin{align}
    \hat{a}_k(t) = \hat{a}_k(0) e^{ - i \omega_k t} - i \sum_m \sqrt{\frac{\omega_k^3 m^2}{32 \hbar L^2 \mu}} \int_{t_0}^t dt' \; \epsilon_{ab} \; \hat{x}^a_m(t') \hat{x}^b_m(t') \; e^{ - i \omega_k (t-t')}.
\end{align}
\end{widetext}
Here $t_0$ is an arbitrary initial time that can be taken to minus infinity if we assume that the environment has a short memory \cite{Borghini2012}. Inserting the solution into the equation of motion (\ref{eq_p}), results in

\begin{widetext}
\begin{align}
     m \Ddot{\hat{x}}_n^j
    &=
    - \sum_{k,m} \frac{m^2 \omega_k^3}{8 L^2 \mu} (\hat{x}_{n})_i \epsilon^{ij}  \int dt'\; \hat{x}^a_m(t') \hat{x}^b_m(t') \epsilon_{ab}  \; \sin(\omega_k(t-t')) + \hat{F}
    \label{eom}\\
    &=  - \sum_{k} \frac{m \omega_k^3}{8 L^2 \mu}  (\hat{x}_{n})_i \epsilon^{ij}  \int dt'  \; \hat{Q}_{ab}(t')  \epsilon^{ab}  \; \sin(\omega(t-t')) + \hat{F}, \label{eomQ}
\end{align}
\end{widetext}
where  $$\hat{F} = \sum_k  \sqrt{\frac{\hbar m^2 \omega_k^3}{8L^2 \mu}} (\hat{x}_{n})_i \epsilon^{ij} (\hat{a}_k(0) e^{-i \omega_k t} + \hat{a}^{\dagger}_k(0) e^{i \omega_k t})$$ is a ``stochastic force", depending on the \emph{initial} creation and annihilation operators of the gravitational environment.
The first term in (\ref{eom}) and (\ref{eomQ}) describes dissipation, i.e. a back reaction force of the particle on the environment after the particle has been accelerated by the interaction with the environment. 

For the quantum description to be an underlying microscopic model for Brownian motion, it is required that the quantum Brownian motion matches the equations for radiation reaction described by general relativity, with a continuous interval of frequencies \cite{Borghini2012}.  For a non-relativistic point particle interacting with gravitational waves, it is known to give a post-Newtonian correction to the metric as a result of the back reaction. To the lowest order, this is the Burke-Thorne potential \cite{burke1971, thorne1969, maggiore}.
$$\Phi(t,\vec{x}) = \frac{G}{5c^5} x^i x^j\frac{d^5Q_{ij}(t)}{dt^5},$$
which generates a force $F_i = -m \partial_i \Phi$. 
This is realized by integrating by parts, where the lower boundary term is suppressed due to the short memory of the environment. The equation (\ref{eomQ}) becomes
\begin{widetext}
\begin{align}
    \ddot{\hat{x}}_n^j
    = -  (\hat{x}_{n})_i   \; [\hat{Q}_{ab}(t)\;  \omega^4_k  + \frac{d^2}{dt^2} \hat{Q}_{ab}(t) \; \omega_k^2 + \frac{d^4}{dt^4} \hat{Q}_{ab}(t)]\; \gamma(0)  \epsilon^{ij}  \epsilon^{ab}
    + (\hat{x}_{n})_i  \; \int dt' \frac{d^5}{dt'^5}\hat{Q}_{ab}(t')   \; \gamma (t-t') \epsilon^{ij}  \epsilon^{ab} + \hat{F},
    \label{EOMQ}
\end{align}
\end{widetext}
where we have introduced the quantity $\gamma(t)$, containing information about the environment and being defined as

\begin{align}
\gamma(t) \equiv  \int d\omega \frac{J(\omega)}{\omega^5} \cos(\omega t).
\label{memkernel}
\end{align}
Here $J(\omega)$ is the spectral density of the environment given by

$$ J(\omega) = \sum_k \frac{\omega_k^3 }{8 L^2\mu} \delta(\omega-\omega_k) = \frac{1}{4 \pi^2 }\sum_{\omega_k} \frac{ G \omega_k^6}{c^5} \; \delta(\omega-\omega_k),$$ where it has been used that  $\frac{1}{L^3} = \frac{|k|^3 }{(2 \pi)^3}= \frac{\omega^3_k}{(2 \pi)^3 c^3}$.

Now, consider the continuous spectral density function $J_c(\omega)$ 
on an interval $\mathcal{I} \equiv [\omega, \omega + d\omega]$, where the width $d\omega$ is much larger than the spacing between frequencies $\omega_k$, although small enough such that $J_c(\omega)$ does not vary significantly over the interval \cite{Borghini2012}. Then we have,
\begin{align}
   J_c(\omega) d\omega = \sum_{\omega_k \in \mathcal{I}} \frac{G  \omega^6}{ c^5},
\end{align}
such that the spectral density $J(\omega)$ in (\ref{memkernel}) can be replaced by the continuous spectrum of bath frequencies $J_c(\omega)$.

The expression for $J_c(\omega)$ on the left side is found by replacing the sum for an integral

\begin{align}
    \frac{1}{L^3} \sum_k \rightarrow \int \frac{d^3k}{(2\pi)^3} = \int \frac{\omega^2}{(2\pi)^3c^3} d\omega d\Omega.
    \label{intdwdo}
\end{align}
Secondly, the polarization tensors are normalized as \cite{toros2024, maggiore} 
\begin{align}
\sum_{\lambda} e_{ij}^{\lambda}(k) e_{kl}^{\lambda}(k)= P_{ik}P_{jl} + P_{il}P_{jk} - P_{ij}P_{kl},
\label{norm_pol}
\end{align}
with the projector $P_{ij} = \delta_{ij}-n_in_j$ and $\vec{n} = \frac{\vec{k}}{|k|}$. To perform the integral over $\Omega$, one can use the identity \cite{maggiore}
\begin{align}
    \int \frac{d\Omega}{4 \pi} \; n_{i_1}\dots n_{i_{2l}} = \frac{1}{(2l+1)!!}(\delta_{i_1 i_2} \delta_{i_3 i_4} \dots \delta_{i_{2l-1} i_{2l}}).
\end{align}

This leads to
\begin{align}
    J_c(\omega) = 
    \begin{cases}
      \frac{2 G}{5 c^5} \;\omega^5 & \text{for}\; 0 \leq \omega \leq \Lambda\\
      0 & \text{otherwise.}
    \end{cases}
\end{align}
Plugging this into (\ref{memkernel}), leads to $\gamma(t-t') = \frac{2G}{5c^5} \delta_{\Lambda} (t-t')$. Here $\delta_{\Lambda}(t-t') = \frac{\sin{(\Lambda (t-t'))}}{(t-t')}$ is a function that is approximately a $\delta(t-t')$ in the limit $\Lambda \rightarrow \infty$ \cite{Borghini2012}, with large values for $(t-t') \in [0,\Lambda^{-1}]$. Hence in the instantaneous time limit the integral in the evolution (\ref{EOMQ}) matches the gravitational radiation force.

Now that we have established the relation to general relativity, we shift our focus to the boundary terms in the equation of motion.
The boundary terms in (\ref{EOMQ}) consist of cut-off dependent terms and diverges when $\Lambda \rightarrow \infty$. The infinities are renormalized by adding counter terms to the initial Hamiltonian. For instance, a mass correction to the bare mass is expected when particles interact with a field. The second derivative term is exactly canceled by adding a mass correction of the form

\begin{align}
    m \rightarrow m + \frac{m^2 \hat{x}^i \hat{x}_i}{8 \mu L^2},
    \label{masscor}
\end{align}
to the Hamiltonian. Secondly, a redefinition of the potential

\begin{align}
    V(x) = V_0(x) + \sum_k \frac{ m^2 \omega_k^2\hat{x}^4}{32\mu L^2},
    \label{repot}
\end{align}
cancels the first term in the equation of motion. This redefined potential is analogous to the known Lamb shift, and similarly appears in interactions with electromagnet fields \cite{barone1991}. Note that the correction is the potential for the harmonic oscillator with mass $\delta m \equiv \frac{m^2\hat{x}^2}{8 \mu L^2}$.

The interesting part, however, is the fourth derivative and is the focus of this article for the following reasons:

\begin{enumerate}
    \item[i]  the fourth derivative exclusively arises when interacting with a gravitational field, and thus signifies a huge difference with open quantum systems interacting with other media. This allows us to examine distinct features related to gravity.
    \item[ii]  Adding lower-derivative corrections, such as the mass correction and potential redefinition described above, will only perturb the original theory. Adding a (unconstrained) higher-derivative correction term, however, will change the new theory dramatically \cite{simon}. This is true even for a small coefficient, i.e., even for a weakly coupled gravitational field, and thus cannot naively be ignored.
    \item[iii] We also emphasize that the Lamb shift is an experimentally verified physical feature, and mass corrections to the bare mass are considered equally real, existing features. Considering that the fourth derivative has the same mathematical origin, it would create a bias if one would disregard the fourth derivative as an unphysical anomaly. The fourth derivative is furthermore of the same order in $G$, the gravitational constant, and thus can also not simply be disregarded as a higher-order contribution in the expansion of $G$.
\end{enumerate}

\section{Higher derivative mechanics}
Besides the cut-off dependency, and divergent behavior in the continuous limit, one may be concerned about other pathologies that are known in higher-derivative theories. For instance, the addition of higher derivatives leads to twice as many solutions for the unitary evolution of the system as for the non-interacting particles \cite{woodard2015,simon}. Half of these solutions are ``run-away" solutions, which come with infinite negative energy modes. However, the lack of a low energy bound is not necessarily manifest in dissipation theories \cite{simon}.

The system is regulated by its small parameter $\frac{G}{c^5}$, and one can consider the perturbative constraint that comes with non-local, low-energy effective theories. The mechanism behind this constraint is to consider the higher derivative solutions as a higher order expansion, and to only allow the solution that reduces to the lower derivative solutions in the appropriate limit \cite{simon}.

We point out though, that the lower derivative theory of the gravitational interaction theory described here, is \emph{not} an effective lower energy theory still describing interaction. The equation of motion shows an expansion around the parameter $\mu$ (or box length $L$). This term in the expansion comes with both lower as well as higher derivative terms. In other words, the mass correction and redefined potential, given by the zeroth and second derivative, are of the same order in $\mu$, as the fourth and fifth derivatives.  Taking the limit $\mu \rightarrow  \infty$ (or $L \rightarrow \infty$) equals the flat spacetime limit, and as such the equations reduce to the \emph{non-interaction theory}, rather than a lower energy \emph{interaction theory}. As such one can not naively describe the gravitational interaction, discuss corrections of order $\frac{G}{c^5}$, and ignore the higher derivative mechanics that come with it. 

The constraints apply to the fifth derivative radiation force. The fourth derivative, however, like the mass correction and potential shift, can be removed with a correction to the Hamiltonian. Hence, even though the systems Hamiltonian contains higher derivatives, the correction terms added are, as a matter of fact, \emph{canceling} the fourth derivative term from the equation, and thus does not create any ``run- away" solution for the full system.

The \emph{full} correction to the  Lagrangian $L_c$, including mass correction and redefined potential, is found  by utilizing the general Euler-Lagrange equation 
\begin{widetext}
$$m \Ddot{x}^j = \frac{d}{dx_j}L_c(x_j, \dot{x}_j, \Ddot{x}_j) - \frac{d}{dt}\frac{d}{d\dot{x}_j} L_c(x_j, \dot{x}_j, \Ddot{x}_j) + \frac{d^2}{dt^2} \frac{d}{d \Ddot{x}_j} L_c(x_j, \dot{x}_j, \Ddot{x}_j),$$
\end{widetext}
 which gives the following Lagrangian in terms of $x_n$, and the quadrupole moment $Q_{ij}$
\begin{widetext}
 \begin{align}
     L_c &= \frac{m^2}{32 \mu L^2} \sum_k \epsilon_{ij} \epsilon_{ab} \left(\omega^2_k x^a x^b x^i x^j - \frac{d (x^a x^b)}{dt}  \frac{d (x^i x^j)}{dt} + \frac{1}{\omega^2_k} \frac{d^2 (x^a x^b)}{dt^2}  \frac{d^2(x^i x^j)}{dt^2}\right)\\
     &= \frac{G}{5\pi c^5} \sum_k \omega_k^3  \left(\omega^2_k Q_{ij}Q^{ij} - \dot{Q}_{ij} \dot{Q}^{ij} + \frac{\ddot{Q}_{ij} \ddot{Q^{ij}}}{\omega^2_k} \right).
    \label{lagrangian}
 \end{align}
 \end{widetext}
In the second line, the polarization tensors are integrated over the solid angle, and the definition of $\mu$ is implemented.
 
Since the associated dynamics is now a fourth-order differential equation, four initial conditions are required, and consequently, four canonical variables are required. With the four canonical variables, the associated Hamiltonian is obtained via the Hamiltonian formulation 
\begin{align}
    H_c = \sum_i^2 P_i X_i - L_c,
\end{align}
where $P_i$ and $X_i$ are the redefined canonical variables. The higher derivative theory is most familiar from Ostrogradsky \cite{woodard2015}, where a similar Lagrangian as in (\ref{lagrangian}) is evaluated. The natural step would then be to adapt Ostrogradsky's choice of canonical variable, albeit in terms of the quadrupole moment:

\begin{align}
    X_{ij}^{(1)} &\equiv  Q_{ij}, \quad P_{ij}^{(1)} \equiv \frac{d L}{d \dot{Q}^{ij}} - \frac{d}{dt}\frac{dL}{d \Ddot{Q}^{ij}},\\
    X_{ij}^{(2)} &\equiv  \dot{Q}_{ij}, \quad P_{ij}^{(2)} \equiv \frac{d L}{d Q^{ij}}.
    \label{conj.coord}
\end{align}
which obey the Poisson equations
\begin{align}
    \{X_{ij}^{(a)}, P_{kl}^{(b)}\} =  \delta_{ab} (\delta_{ik} \delta_{jl}+\delta_{il} \delta_{jk}),
\end{align}
and their quantum analog
\begin{align}
     [\hat{X}_{ij}^{(a)}, \hat{P}_{kl}^{(b)}] =  i \hbar \delta_{ab} (\delta_{ik} \delta_{jl}+\delta_{il} \delta_{jk}).
     \label{commrel}
\end{align}
For the commutation relations, the quadrupole moment has been promoted to a quantum operator $\hat{Q}_{ij}$. 

The choice of canonical variables in terms of the quadrupole moment is motivated by the fact that the interaction with the gravitational degrees of freedom is via the quadrupole moment, i.e., $H_{int} \propto q \epsilon^{ij} Q_{ij}$. Furthermore, gravitational radiation is produced by changes in the quadrupole moment \cite{maggiore}. Hence, we are interested in the dynamics of the quadrupole variables, and the higher derivative theory takes a significantly simpler form when described in terms of the quadrupole moment. It is furthermore interesting to consider decoherence and noise emerging in different measurements, such as electric quadrupole measurements. 

As an example, we could consider a nuclear mass distribution,
$R = R_0 A^{\frac{1}{3}}$, instead of a sum of infinitesimal point particles.
Here, $R_0$ is the radius, and the number of
nucleons in the nucleus is given by the atomic number $A$ \cite{Neugart2006}. 
It is expected that the deformation of the nucleus due to the gravitational wave interaction, will produce a dynamical mass quadrupole moment, and affect the electric quadrupole moment as well. As such, signatures of gravitational radiation could in principle be observed in the hyperfine splitting of the nucleus. The idea for searching for gravitational waves in hyperfine splitting, albeit via a different approach than described here, was also suggested by Wanwieng et al. \cite{Wanwieng_2023} for supermassive black hole binaries.

The spectroscopic quadrupole moment is given by \cite{Neugart2006}
\begin{align}
    Q_{ij} = \frac{3 K^2 - I(I + 1)}{(I + 1)(2I + 3)}  Q_{ij}^0,
\end{align}
 with $K$ the projection of the nuclear spin on the deformation axis, and $I$ the nuclear spin. The intrinsic quadrupole moment $Q_{ij}^0$ emerges from the deformation of the spherical charge distribution of protons and is given by \cite{Neugart2006}

 \begin{align}
     Q_{ij}^0 = \frac{3}{\sqrt{5 \pi}} (Z \beta \;x_i x_j +  0.36 \; \beta \; \delta_{ij}).
 \end{align}
Here, $\beta$ is the nucleus charge deformation parameter, which is related to the amplitude of the gravitational wave, via the mean square radius of a deformed nucleus:
\begin{align}
    \expval{x_i x_j} =  \frac{\int dx \; \rho(x,t) x_i x_j}{\int dx \; \rho(x,t)}.
\end{align}
Here, one can consider noise in the time evolution of the quadrupole moment (mass density).

In either case, note that the mass quadrupole moment (\ref{quad}) is independent of the individual particle position variable, and in the case of gravitational waves, a solely time-dependent variable. Hence, the higher derivative commutation relations, defined in (\ref{commrel}) do not necessarily apply to the particle position and momentum operator. In other words, from the commutation relation, we see that  $[\hat{Q}_{ij}, \dot{\hat{Q}}_{ij}]=0$, although it is still considered that $[\hat{x},\hat{p}]= i \hbar$. 

With the redefined canonical variables, the correction Hamiltonian is given by
\begin{widetext}
\begin{align}
    \hat{H}_c =& \frac{G}{5\pi c^5} \sum_k \omega_k^3 (
    - \dot{\hat{Q}}_{ij} \dot{\hat{Q}}^{ij} - \omega_k^2 \hat{Q}_{ij} \hat{Q}^{ij}  + \frac{\ddot{\hat{Q}}_{ij} \ddot{\hat{Q}}^{ij}}{\omega_k^2} - \frac{2}{\omega_k^2}\; \dot{\hat{Q}}_{ij} \dddot{\hat{Q}}^{ij}) \label{H_C}
    \\
    =&  \hat{P}_{ij}^{(1)} \hat{X}^{ij}_{(2)} + \sum_k \frac{5 \pi c^5}{2 G \omega_k}\; \hat{P}_{ij}^{(2)} \hat{P}^{ij}_{(2)} +   \frac{G}{5\pi c^5} \sum_k \omega_k^3 ( \hat{X}_{ij}^{(2)} \hat{X}^{ij}_{(2)} - \omega_k^2 \hat{X}_{ij}^{(1)} \hat{X}^{ij}_{(1)}).
\end{align}
\end{widetext}
The first two terms in (\ref{H_C}) are the kinetic part and potential of a harmonic oscillator for $\hat{Q}$ that is associated with the mass correction (\ref{masscor}) and redefined potential (\ref{repot}), respectively. The last two terms are the result of the fourth-order derivative in the equation of motion.

The associated Euler-Lagrange equation for $Q_{ij}$ is given by
\begin{widetext}
\begin{align}
    F_{ij} &= \frac{d}{dQ^{ij}} L_c(Q_{ij},\dot{Q}_{ij}, \ddot{Q}_{ij}) - \frac{d}{dt} \frac{d}{d\dot{Q}^{ij}} L_c(Q_{ij},\dot{Q}_{ij}, \ddot{Q}_{ij}) +\frac{d^2}{dt^2} \frac{d}{d\ddot{Q}^{ij}} L_c(Q_{ij},\dot{Q}_{ij}, \ddot{Q}_{ij})\\
    &=\frac{G}{5\pi c^5} \sum_k \omega_k^3  (\omega_k^2 Q_{ij} +\ddot{Q}_{ij}  + \frac{\ddddot{Q}_{ij} }{\omega_k^2}),\label{eomforQ}
\end{align}
\end{widetext}
where the dynamics for a single particle given by (\ref{eomQ}) is rediscovered by $F^j =  \frac{d Q_{ij}}{dx_j} F^{ij}$.
The solution of the Euler-Lagrange equation (\ref{eomforQ}) is
\begin{widetext}
\begin{align}
    Q_{ij}(t) = Q^+_{ij}(0)\cos(\omega_+ t) + \frac{\dot{Q}^+_{ij}(0) }{\omega_+}\sin(\omega_+ t) +Q^-_{ij}(0)  \cos(\omega_- t) +\frac{ \dot{Q}^-_{ij}(0) }{\omega_-}\sin(\omega_- t)
    \label{soln}
\end{align}
with the initial conditions
\begin{align}
    Q^+_{ij}(0) = \frac{\Ddot{Q}_{ij}(0) + \omega_-^2 Q_{ij}(0)}{\omega_-^2 - \omega_+^2}, \quad Q^-_{ij}(0) = \frac{\ddot{Q}_{ij}(0) + \omega_+^2 Q_{ij}(0)}{\omega_+^2 - \omega_-^2},
    \label{initcond}
\end{align}
\end{widetext}
and frequencies
\begin{align}
    \omega_{k \pm} \equiv \frac{\omega_k}{2} (1 \pm i\sqrt{3}) = \omega_k \; e^{\pm \frac{ i \pi}{6}}. \label{freq}
\end{align}
Note that the quadrupole moment evolves with the \emph{gravitational frequencies} which have been split into positive and negative imaginary parts. Hence, we define the gravitational operators in terms of creation operators for the excitations with frequencies $\omega_{k_+}$ and $\omega_{k_-}$,
\begin{align}
    \hat{q}_l&= \sum_{l = k_+,k_-}\;  \text{sign}(l) \; \sqrt{\frac{\hbar}{2 \mu  \omega_l}}(a_l(t) + a^{\dagger}_l(t)) \\
    \dot{\hat{q}}_l&= i \sum_{l = k_+,k_-}\;  \text{sign}(l) \; \sqrt{\frac{\hbar \omega_l}{2 \mu }}(a_l(t) - a^{\dagger}_l(t)) \\
    \ddot{\hat{q}}_l&= \sum_{l = k_+,k_-}\;  \text{sign}(l) \;  \omega_l \sqrt{\frac{\hbar  \omega_l}{2 \mu }}(a_l(t) + a^{\dagger}_l(t)) \\
    \dddot{\hat{q}_l} &= i \sum_{l = k_+,k_-}\;  \text{sign}(l) \; \omega_l^2 \sqrt{\frac{\hbar \omega_l}{2 \mu }}(a_l(t) - a^{\dagger}_l(t)),
\end{align}
with sign$(l)$ being the sign of the imaginary frequency (i.e. sign$(\Im(\omega_{k_{\pm}}))$).
It is considered that the excitations with positive and negative energy modes are distinct excitations (see \cite{Ganz_2021}), and as such the two pairs of annihilation/creation operators commute with each other: $[\hat{a}_{k_{\pm}}, \hat{a}^{\dagger}_{k_{\mp}}] = 0$.

\section{Quantum master equation}

In the last section of this article, we shall derive the quantum master equations for the higher derivative model described in the previous sections. For this, we consider the reduced density matrix 
\begin{align}
     \hat{\rho}^I_s(t) = -\frac{1}{\hbar^2} \Tr_{q}\int_0^t dt' \int_0^{t'} dt''\; [\hat{H}^I_{int}(t'),[\hat{H}^I_{int}(t''), \hat{\rho}^I(t) ]],
\end{align}
where the superscript $I$  indicates the interaction picture. This followed from iterative integration of the Liouville–von Neumann equation,
$$\frac{d \hat{\rho}^I(t')}{dt'} = -\frac{i}{\hbar} [\hat{H}^I_{int}(t'), \hat{\rho}^I(t') ],$$ with on the right side $$\hat{\rho}^I(t') = \hat{\rho}(0) - \frac{i}{\hbar} \int_0^{t'} dt'' [\hat{H}^I_{int}(t''), \hat{\rho}^I(t'') ].$$ Then both sides are integrated over $t'$, where the lower boundary can be ignored due to the Markov approximation. For the same reason, we have $\hat{\rho}(t'') \approx \hat{\rho}(t')\approx \hat{\rho}(t)$ and we assume that he system and environment were initially uncorrelated such that $[\hat{H}^I_{int}(t'), \hat{\rho}(0) ]=0$.

The interaction Hamiltonian is $\hat{H}_{int} \propto \ddot{\hat{q}}_k \; \hat{Q}_{ij}\epsilon^{ij}= \omega^2_k \hat{q}_k\; \hat{Q}_{ij} \epsilon^{ij}$. The trace over the environmental degrees of freedom will give the self-correlation functions 
\begin{widetext}
\begin{align}
    \omega_{k+}^2 \omega_{k-}^2 \expval{\hat{q}_k(t'') \;\hat{q}_k(t')}=& \sum_{l = k+,k-} \frac{ \hbar \omega_{k+}^2  \omega_{k-}^2}{2  \mu\omega_l} \left( (2 N_{l} -1)\cos(\omega_{l} (t''-t')) + i\; \text{sign}(l) \sin(\omega_{l}(t''-t'))\right) \label{corrq}\\
     \expval{\ddot{\hat{q}}_k(t'') \; \ddot{\hat{q}}_k(t')}=& \sum_{l = k+,k-}  \frac{\hbar  \omega_l^3}{ 2 \mu} \left( (2 N_{l} -1)\cos(\omega_{l} (t''-t')) + i\;\text{sign}(l) \sin(\omega_{l}(t''-t'))\right)\label{corrddotq}
\end{align}
\end{widetext}
where  $N_{\pm} = \expval{\hat{a}_{k_{\pm}}^{\dagger}\; \hat{a}_{k_{\pm}}}$ is the averaged number of gravitons, and from (\ref{freq}) we see that $\omega_k^4 = \omega_{k-}^2 \omega_{k+}^2$. The two correlation functions are in principle equal with a slightly different prefactor in terms of $\omega_{k_{\pm}}$. The noise and dissipation kernel $\nu_{\pm}(\tau)$ and $\eta_{\pm}(\tau)$ are then given by

\begin{align}
    \nu_{\pm}(t''-t') &=\frac{1}{\hbar}  \int_0^{\infty} d \omega J(\omega_{\pm}) (2 N_{\pm} -1)\cos(\omega_{\pm} (t''-t')) \nonumber\\
    i \eta_{\pm} (t''-t') &= \pm  \frac{i}{\hbar} \int_0^{\infty} d \omega J(\omega_{\pm}) \sin(\omega_{\pm}(t''-t')),\nonumber
\end{align}
with a spectral density $J(\omega_+) = \sum_k \frac{ \omega_+ \omega_-^2}{32 L^2 \mu} \delta(\omega - \omega_{k})$.  

For the quadrupole moment, one could insert the solution given by (\ref{soln}, or Taylor expand $\hat{Q}$ till the third order and integrate out the $t^n$ via partial integration. With an eye on the determined correction Hamiltonian, we deduce that the part solely containing lower derivatives of the quadrupole moment interacts with the self-correlation function (\ref{corrq}), while any part containing at least one higher derivative of $\hat{Q}$ interacts with (\ref{corrddotq}). This marks a difference with lower derivative systems where only the correlations between the amplitude $\hat{q}(t)$ plays a role. Here it is suggested that both the correlations between the amplitude as well as the acceleration affect the system. One can easily check that the imaginary part of the master equation is consistent with the derived correction Hamiltonian by multiplying the correlation functions with an overall factor of $\frac{-\omega_{k+}^2 \omega_{k-}^2}{(\omega_{k-}^2 - \omega_{k+}^2)^2}$. Then

\begin{widetext}
\begin{align}
  \Im \left( \frac{ d\hat{\rho}_s(t)}{d t}\right)
     =& -  \frac{G}{5\pi \hbar c^5} \sum_k \omega_k^3 \Big[ \frac{-\ddot{\hat{Q}}^2}{\omega_{k-}^2-\omega_{k+}^2}+\frac{\omega_{k+}^2 \omega_{k-}^2 \hat{Q}^2  }{\omega_{k-}^2-\omega_ {k+}^2} + \frac{(\omega_{k-}^2 + \omega_{k+}^2) \dot{\hat{Q}}^2}{\omega_{k-}^2-\omega_{k+}^2} + \frac{2 \dot{\hat{Q}} \dddot{\hat{Q}}}{\omega_{k-}^2-\omega_{k+}^2}, \hat{\rho}_s(t)\Big]\\
    &+\frac{1}{\hbar} \sum_{l= +, -} \int_0^t dt' \gamma_l(t'-t) \;[\ddot{\hat{Q}} \dddot{\hat{Q}} +\dddot{\hat{Q}} \ddot{\hat{Q}}, \hat{\rho}_s(t)],
    \label{imrho}
\end{align}
\end{widetext}
with $ \gamma_l (t'-t) = \int d \omega  \frac{ J(\omega_l)}{ \omega_l (\omega_+^2 - \omega_-^2)^2}  \cos(\omega_l(t'-t))$, and the boundary term is exactly canceled by $H_c$. One can once again replace the spectral density $J(\omega)$ by $J_c(\omega)$, with
\begin{align}
    J_c(\omega) = 
    \begin{cases}
      \gamma_0 \; \omega_+ ( \omega_+^2 - \omega_-^2)^2 & \text{for}\; 0 \leq \omega \leq \Lambda\\
      0 & \text{otherwise.}
    \end{cases}
\end{align}
which  results in $\gamma(t'-t) =  \gamma_0 \;  \delta_{\Lambda}(t'-t)$, with $\delta_{\Lambda}(t'-t) =  \frac{\sin{(\Lambda (t'-t))}}{t'-t}$. The last term in (\ref{imrho}) is dependent on $\ddot{\hat{Q}} \dddot{\hat{Q}}$, which is recognized  from general relativity as the loss of momentum through radiation \cite{maggiore}, 
\begin{align}
    \frac{dJ}{dt} = \frac{2G}{5 c^5} \expval{\ddot{Q}_{ij} \dddot{Q}^{ij}}.
\end{align}
Here, $J$ is the spin and angular momentum for quantum particles. The  radiated energy, proportional to $\expval{\dddot{Q}_{ij} \dddot{Q}^{ij}}$, is of order $\order{\frac{1}{\Lambda}}$  and can be neglected in the limit $\Lambda \rightarrow \infty$.

The leading order of the real part  is given by

\begin{widetext}
\begin{align}
    \Re \left(\frac{d\hat{\rho}_s (t)}{dt} \right) 
=\; \frac{G}{5 \pi \hbar c^5} \sum_k \omega_k^3  \int_0^t  dt' & \Bigg(\frac{ \omega_{k_+}^3 \omega_{k_-}^4}{(\omega_{k_+}^2-\omega_{k_-}^2)^2}  (2 N_+ - 1)  \; \cos{\omega_{k_+}(t-t')} \nonumber\\
&+ \frac{ \omega_{k_-}^3 \omega_{k_+}^4}{(\omega_{k_+}^2-\omega_{k_-}^2)^2}  (2 N_- - 1)  \; \cos{\omega_{k_-}(t-t')}\Bigg) [\hat{Q}_{ij}, [\hat{Q}^{ij} ,\hat{\rho}_s(t)]].
\label{lead}
\end{align}
\end{widetext}
We will consider the environmental vacuum state ($N=0$). The Lindblad form is found by writing the quadrupole moment as
\begin{align}
    \hat{Q}_{ij}(t) = \hat{X}_{ij}^{(1)} = \frac{\hbar}{\omega_Q} (\hat{b}_{ij}\; e^{i\omega_Q t} + \hat{b}_{ij}^{\dagger} e^{-i\omega_Q t}),
    \label{quadladder}
\end{align}
with $\omega_Q$ the quadrupole frequency for which $\omega_Q = \omega_k$ hold, if we consider that the quadrupole is entirely generated by the gravitational waves. Here, we apply the rotating wave approximation \cite{toros2024, Agarwal_2012}, such that only terms with $~ \hat{b}\hat{b}^{\dagger}$ remain. Secondly, we write the cosine in its exponential form and define $\tau\equiv t-t'$. Note that the short memory of the environment allows us to take the upper integration boundary to infinity. Then with
\begin{align}
    \int_0^{\infty} d\tau \; e^{ - i (\omega_Q - \omega_{k_{\pm}}) \tau} = \pi \delta(\omega_Q - \omega_{k_{\pm}}),
\end{align}
where we impose that the real part of the frequencies must be positive, we find:
\begin{align}
    \Re \left(\frac{d\hat{\rho}_s (t)}{dt} \right) 
=&  \frac{G \hbar}{15 c^5} \omega_Q^3\; (\hat{b}^{ij} \hat{\rho}_s \hat{b}_{ij}^{\dagger} - \frac{1}{2}\{  \hat{b}^{ij} \hat{b}_{ij}^{\dagger} , \hat{\rho}_s \} ), \nonumber
\end{align}
with $\frac{G \hbar }{c^5} = t_p^2$, the Planck time squared. The decoherence rate is consistent with previous results \cite{toros2024, Oniga2017}, although with a different decoherence operator. The operator $\hat{b}_{ij}$ is the mode operator for quadrupole excitations with frequency $\omega_Q = \omega_k$. A significant difference with lower derivative models is that the system is no longer described by ``regular" particle position and momentum operators. 

In canonical variable, the equation has the form

\begin{align}
    \frac{d\hat{\rho}_s (t)}{dt} = D [\hat{X}^{(1)}_{ij}, [\hat{X}_{(1)}^{ij}, \hat{\rho}_s(t)]],
\end{align}
with $D$ the diffusion coefficient. This shows decoherence in the canonical momentum $\hat{P}_{(1)}$ basis, where $\hat{P}_{(1)} \propto \dot{\hat{Q}} + \frac{\dddot{\hat{Q}}}{\omega_k^2}$. One could have chosen to define the canonical variables in terms of individual particle coordinates. Then the diffusion equation would take the familiar form $\frac{d\hat{\rho}_s (t)}{dt} = D [\hat{X}^2_1, [\hat{X}^2_1, \hat{\rho}_s(t)]],$ with $X_1 \equiv x_n$. The associated canonical momentum is then given by a complex combination of position, velocity, acceleration, and jerk of the particle.

It is also interesting to look at the next-to-leading order term which is given by
\begin{widetext}
\begin{align}
    \Re \left(\frac{d\hat{\rho}_s (t)}{dt} \right) 
=&  \int d t' \frac{-G  \pi \omega_{k_-}^2 }{ \hbar c^2 L^3} \; \cos{\omega_{k_+}(t-t')} \Bigg( \left[\frac{\Ddot{\hat{Q}}_{ij}}{\omega_{k_-}^2}, [\dot{\hat{Q}}^{ij} ,\hat{\rho}_s(t) ]\right] + [ Q_{ij}, \left[\dot{\hat{Q}}^{ij} + \frac{\dddot{\hat{Q}}^{ij}}{\omega_{k_-}^2}, \hat{\rho}_s(t) \right]]\Bigg) \nonumber \\
=& \int d t' \; \int d\omega \frac{J(\omega_{+}) }{\omega_{+} \hbar} \;  \cos{\omega_+(t-t')}  \left( [\hat{X}_{ij}^{(1)}, [\hat{P}_{(1)}^{ij}, \hat{\rho}_s(t)]] - [\hat{P}_{ij}^{(2)}, [\hat{X}_{(2)}^{ij}, \hat{\rho}_s(t)]] \right). \label{rerho}
\end{align}
\end{widetext}
In terms of canonical variables, we can recognize this part as anomalous diffusion. By solving the integrals, one can see that the term in the continuous spectrum will be of order $\order{\Lambda^4}$. Hence, the anomalous diffusion can not be ignored based on an irrelevance argument in the limit $\Lambda \rightarrow \infty$.

\section{Summary and discussion}

In this article, we have re-evaluated gravitational radiation from a system interacting with gravitational waves. We have shown that gravitational waves generate a mass quadrupole moment in a system of free particles, where the quadrupole moment naturally evolves according to higher derivative dynamics. The result is that along with physical features such as mass correction and frequency shifts, a higher derivative feature is expected in the unitary description of the quadruple moment. All these terms are of leading order in the expansion of the gravitational constant $G$. This has significant implications for the model since the unitary part of the system now contains a larger set of solutions than before the gravitational interaction. In the full theory, this extra higher derivative term cancels a fourth derivative in the Langevin equation that emerges from the interaction Hamiltonian. Hence, unlike the radiation force, which possibly needs constraints, the fourth derivative does not create instability issues. 
The most noteworthy implication is the larger number of required independent operators, which has implications for the basis in which decoherence is expected. We found that the system decoheres in the canonical momentum $P_{(1)}$ basis, with $P_{(1)}$ consisting of velocity and jerk of the quadrupole moment operator. One could define the canonical variable in terms of particle position $x_n$ and velocity $\dot{x}_n$. This would, however, still lead to decoherence in a canonical momentum basis, where the canonical momentum depends on position, momentum, acceleration, and the jerk of the particle. The fourth derivative is furthermore interesting since it marks a significant difference with systems interacting with other media.

For the canonical operator, we have defined the operators as quadrupole operators and continued focusing on the dynamics of the quadrupole moment, rather than the individual particles. This was motivated by several reasons, First, the radiation friction, as well as the corrections to the Hamiltonian, were all written in terms of the quadrupole. The higher derivative model also took a simpler form when regarding the quadrupole moment. 

Secondly, we gave a short example of a potential method to probe gravitational decoherence and noise in quadrupole dynamics. Here the mass quadrupole is related to the electric quadrupole, and it is expected that gravitational decoherence and noise will translate to the electric quadrupole moment. The electric quadrupole can be measured via e.g. M\"{o}ssbauer effect to very high precision. This requires more research and is reserved for future work.

Lastly, we have derived a quantum master equation and deduced that the self-correlation function between the amplitude $\hat{q}$ as well as the acceleration $\ddot{\hat{q}}$ are essential for reproducing the result obtained from the study of the Langevin equation.
The decoherence rate is consistent with earlier results in the literature, although the basis in which decoherence is expected is in a canonical momentum basis, as mentioned above. This marks a significant difference with lower derivative models, where decoherence is always considered in the usual particle position or momentum basis.

\section*{Acknowledgments}	
	
	The author would like to thank A. Gro\ss ardt for helpful discussion, and thank H. Gies and A. Gro\ss ardt for comments on this manuscript. The author gratefully acknowledges the support of the Volkswagen Foundation, and the Deutsche Forschungsgemeinschaft (DFG) under Grant No 406116891 within the Research Training Group RTG 2522/1.

\bibliography{main} 
\bibliographystyle{ieeetr}

\end{document}